\documentclass[a4paper]{VisionStyle}
\usepackage{epsfig}
\def\c{{\sl Chandra}}

\def\scuba{{\sl SCUBA}}
\def\zp{$z_{\rm phot}$}

\def\j{{\sl J}}
\def\h{{\sl H}}
\def\k{{\sl K}}
\def\b{{\sl B}}
\def\i{{\sl I}}
\def\u{{\sl U}}

\def\ltsim{\mathrel{\hbox{\rlap{\hbox{\lower4pt\hbox{$\sim$}}}\hbox{$<$}}}}
\def\gtsim{\mathrel{\hbox{\rlap{\hbox{\lower4pt\hbox{$\sim$}}}\hbox{$>$}}}}

\def\nh{N$_{\rm H}$}

\def\ergpspsqcm{ erg s$^{-1}$ cm$^{-2}$}
\def\ergps{ erg s$^{-1}$}

\def\xmm{{\sl XMM-Newton}}
\def\micron{${\rm{\mu}m}$}

\def\int{{\sl INT}}

\def\scuba{{\sl SCUBA}}
\def\sirtf{{\sl SIRTF}}
\def\zp{$z_{\rm{phot}}$}
\def\zs{$z_{\rm{spec}}$}

\def\ha{H$\alpha$}
\def\hb{H$\beta$}

\def\pab{Pa$\beta$}
\def\oiii{[OIII]}

\def\nii{[NII]}
\def\sii{[SII]}

\def\l{$\lambda$}
\def\oi{[OI]}

\begin{document}

\title{Powerful Obscured AGN in \c\ cluster fields}

\author{P.\,Gandhi \and A.C.\,Fabian \and C.S.\,Crawford } 

\institute{
  Institute of Astronomy, Madingley Road, Cambridge CB3 0HA, United Kingdom }

\maketitle 

\begin{abstract}

Models for the origin of the hard X-ray background have suggested that sources with the most accretion activity lie hidden in highly obscured AGN. We report on our study of hard, serendipitous sources in the fields of \c\ clusters with fluxes close to the turn-over in the source-counts. These include two Type II quasars with measured X-ray luminosities $>10^{45}$ \ergps\ and column-densities $> 10^{23}$ cm$^{-2}$, one possibly being Compton-thick. Both show indications of redshifted Fe K$\alpha$ line emission. Radiative transfer modelling of the broad-band spectrum of a highly-magnified source with deep {\sl ISOCAM} detections implies the presence of warm-to-hot dust obscuring a quasar with L$_{\rm UV}>10^{45}$~\ergps. Multi-wavelength spectroscopic and photometric follow-up of the optically-faint sources suggests that these objects are found in the centres of massive evolved galaxies at a range of redshifts, with red optical / near-infrared colours dominated by the host galaxy. Detailed source identification is difficult due to the paucity of strong emission features, especially in the near-infrared. We present the main results from a sample of near-infrared spectra of optically-faint sources obtained with 4~m and 8~m telescopes. Through the study of the harder and brighter X-ray background source population, we are likely to be viewing the most intense phase of the growth of supermassive black holes.

\keywords{
X-rays: diffuse background -- 
X-rays: galaxies -- 
infrared: galaxies -- 
galaxies: active --
quasars: Type II
}
\end{abstract}

\section{Introduction}

The \c\ observatory has largely resolved the hard (2--10 keV) X-ray
background (HXRB) within two years of its launch, after almost four
decades of scientific effort (\cite{giacconi}; \cite{mushotzky}). Follow-up work within the past year has revealed that this \c\
source population can be broadly split into quasars (a small fraction), narrow emission-line AGN, optically-normal galaxies
with no sign of activity other than in X-rays and optically-faint
sources which are difficult to identify (\cite{bargeretal}; \cite{giacconi2001} \cite{alexanderfaint}; \cite{willott}). {\sl XMM-Newton}, with its higher effective area at 10 keV, has
also begun to deliver results (\cite{hasinger}; \cite{barconsAXIS}) and the essential \c\ findings are confirmed with many new Type II AGN being found (eg, \cite{lehmannxmm}).

The individual X-ray spectra of the hard sources are flat enough to account for the HXRB
spectral slope of 0.4 (\cite{marshall}; \cite{gruber}), and
the integrated flux can contribute between $\sim 70-105$\% of the HXRB intensity. The ambiguity in the absolute known intensity is still to be resolved. Its cause is cross-calibration mismatches between various X-ray missions and/or cosmic variance of the sources themselves (eg, \cite{barcons}; \cite{cowie_number_counts}). Essentially, these observations bear out the main prediction of models which
synthesize the HXRB through the integrated emission of a population of
obscured active galaxies (AGN) with a distribution of absorbing
columns and spread over redshift (\cite{setti}; \cite{comastri}, \cite{wfn}). The hard spectral slope is produced by the presence of large obscuring columns which erode the soft photons. This also has the exciting corollary that \lq Type II\rq\ QSOs -- the missing population of powerful, absorbed sources should emerge in large numbers.

In fact, the optically-faint sources are prime candidates for being members of this Type II population. Due to their faintness, it has been difficult to determine their nature even with 10 m class telescopes (eg, \cite{bargeretal}). Speculations include the possibility that these are very high redshift ($>$6)
quasars, though \cite*{alexanderfaint} find that such objects can be at most a small fraction of the total population. 

We have been studying the
optically-faint population of hard sources in \c\ cluster fields. The main aim-point of the observation is the cluster itself and typical exposure times range from 10--40~ks. We thus have $0.5-7$~keV flux coverage for point background sources of $3\times 10^{-15}\sim 10^{-13}$~\ergpspsqcm. The contribution to the XRB is maximized at a turn-over in the source counts at $\approx 10^{-14}$~\ergpspsqcm\ and we thus study the dominant flux contributors to the XRB by targeting this regime (eg, \cite{cowie_number_counts}). It is likely that $40\sim 70\%$ of the 2--10~keV XRB is emitted from sources at such fluxes.

Unlike very deep, small-area surveys, our minimal selection criterion is absence or faintness on the Digitized Sky Survey\footnote{http://archive.stsci.edu/dss/}. In this way we are able to select hard X-ray sources with weak optical counterparts from almost anywhere in the sky. We find that many of these sources are relatively bright and readily detected in the near-infrared (NIR), with \k$\sim 17$ (\cite{c01a}, \cite{moriond}). The large optical--IR colours (reaching as much as  $R-K\approx 8$; see also \cite{alexanderred}) are consistent with the flux being dominated by
light from the host galaxy and any central AGN being highly obscured.
Photometric redshift (\zp; \cite{hyperz}) estimates suggest that these are early-type galaxies at redshifts $z\sim 1-2$, with a tail of sources at higher redshifts. One such highly obscured source was confirmed to be a Type~II QSO in
the field of Abell~2390 (\cite{c01b} and also discussed in section 2 of this work).

Herein we report the discovery of another Type II QSO in the field of the Abell~963 cluster. We also find potential redshifted Fe K$\alpha$ lines in both the above Type~II QSOs. In addition, we report on our findings of near-infrared spectra of hard serendipitous \c\ sources. Through 4~m and 8~m telescope observations, we have found that few sources contain strong detectable emission lines and fewer still can be classified with confidence. This is most likely due to depletion / destruction of emission lines by large column-density gas and the associated dust. We also report on indications of clustering of hard \c\ sources in the field of A\,963. 

Quoted cosmological quantities assume H$_0=50$ km~s$^{-1}$ Mpc$^{-1}$ and q$_0=0.5$ throughout.

\section{The field of Abell~2390}

The \object{cluster Abell~2390} ($z=0.228$) was observed with \c\ on 2000 October 08 and 1999 November 05 for a total exposure time of 19 ks. The emission from the cluster was concentrated within a few arcmin of the ACIS-S3 background-illuminated chip and we were able to extract 31 sources in the full $0.5-7$ keV band (\cite{c01b}) from the rest of the ACIS-S array. Roughly one-third (10) of these were found to have a hard X-ray count ratio, i.e. a soft(0.5--2; S):hard(2--7; H) band ratio $<2$, or hardness ratio (H-S)/(H+S)$>-0.3$. This is approximately the ratio that is predicted to be observed on the ACIS-S3 chip for an AGN at $z=2$ emitting power-law radiation (with photon-index $\Gamma=2$) obscured by gas with column density 10$^{23}$~cm$^{-2}$ at the redshift of the source, while the Galactic column in the line-of-sight toward A\,2390 is $7\times 10^{20}$ cm$^{-2}$ only. Thus, these hard sources are consistent with being highly obscured AGN. 

Using multi-wavelength optical-infrared photometry to estimate photometric redshifts and Keck optical spectroscopy, we find that 17 sources lie at a large range of redshifts $(0.2-3)$ with a median $z$ of 1.47. Most sources are consistent with Compton-thin Seyfert IIs. There are 2 narrow-line AGN; 1 unobscured AGN with broad emission lines and blue optical colours; and 1 AGN with broad lines and blue optical/NIR colours but high ($\approx 3\times 10^{22}$~cm$^{-2}$) obscuration to its nucleus inferred from X-rays, implying a dust:gas ratio different from the Galactic value (cf. \cite{maiolino}; \cite{wilkes2mass}). Two sources in particular have X-ray $2-10$~keV luminosities of $2\times 10^{45}$ and $2\times 10^{44}$ \ergps\ respectively, after correction for strong gravitational lensing by the cluster and the intrinsic obscuring column densities ($\approx 2\times 10^{23}$~cm$^{-2}$) measured from the X-ray spectra. These are hereafter referred to as source A and source B respectively. Thus, source A is a genuine X-ray Type II QSO ($z=1.467$; \cite{cowie}). 

\begin{figure}[ht]
  \begin{center}
    \epsfig{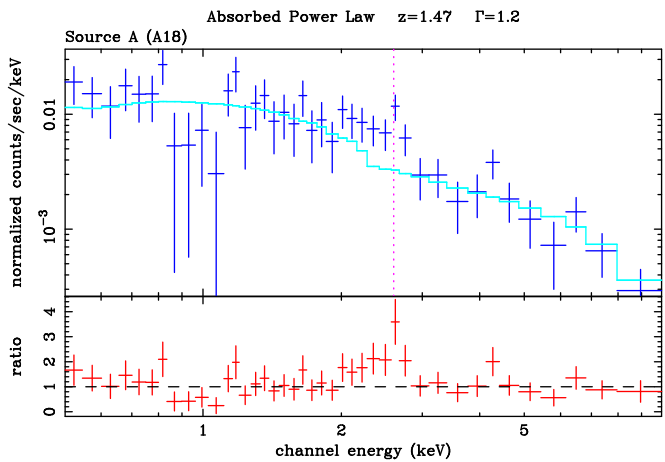}
  \end{center}
\caption{ \xmm\ PN extracted spectrum for the Type~II QSO (source A) in the field of Abell~2390. The data is shown as the dark blue crosses with 1$\sigma$ errors. The model (light-blue line) is a $\Gamma=1.2$ power-law transmission model at $z=1.467$ affected by Galactic absorption. The ratio plot (of the data to the model; below) shows that there is an excess of counts at $\approx 2.6$~keV, exactly the energy where a redshifted Fe K$\alpha$ line is expected to lie (pink dotted line).}
\label{fig:a18xmmpn}
\end{figure}

We recently obtained \xmm\ data for the field of A\,2390 (effective exposure time 18 ks), and extracted a spectrum of source A from the PN chip (Fig~\ref{fig:a18xmmpn}). A power-law transmission model at $z=1.467$ fitted to the data implies a hard spectrum, with a free-fitting $\Gamma=1.2$. In addition, there is an excess of counts at exactly the energy where a redshifted 6.4 keV (rest-frame) neutral Fe K$\alpha$ line is expected to lie. A simple narrow redshifted gaussian line fit to this excess implies a rest-frame equivalent-width $\approx 1$ keV. However, this source lies only $\approx$~1.2~arcmin from the cluster core and, owing to the relatively large point spread function of \xmm, the background is dominated by cluster emission and its accurate removal is difficult. Thus, without more target photons (the source has $\approx 570$ counts in the total band), it is difficult to constrain the parameters of the Fe line further. We note that the non-detection of any obvious Fe line in the \c\ spectrum (\cite{c01b}) is completely consistent with the fewer counts observed ($\approx 70$ source counts in each of two ACIS exposures).
 
Both the luminous sources A and B were also detected in an ultra-deep {\sl ISOCAM} exposure of the A\,2390 field (\cite{altieri}). Source B is magnified by a factor of 8 and its detections were used by \cite*{wfg} -- see also \cite*{c01b} -- to perform radiative transfer modelling of the broad-band spectrum. Assuming power-law radiation incident on an obscuring sphere of gas mixed with dust (in the typical ISM ratio), models were produced for various parametric values of the optical depth, temperature and spatial scale of the surrounding dust. It was found that the significant {\sl ISOCAM} detections (\cite{lemonon}) combined with a deep \scuba\ 850\micron\ upper-limit (\cite{f00}) could be fit with warm-to-hot (as much as 1500 K in the innermost regions) dust with optical-depth in agreement with that inferred from the X-ray gas column-density measurements (assuming a Galactic dust:gas ratio) and extending from $0.1\sim 50$~parsec from the nucleus. The implied absorbed nuclear luminosity (modelling the big blue bump) is $\approx 3\times 10^{45}$ \ergps\ (at \zp=2.78; this power has been corrected for lensing), suggesting that more such luminous absorbed QSOs should be found with the upcoming \sirtf\ mission.

\section{The field of Abell~963}

The 36~ks ACIS exposure of the \object{cluster A\,963} (2000 November 10) yielded 60 serendipitous sources with WAVDETECT\footnote{http://cxc.harvard.edu/ciao/}. Again, roughly one-third (18) had S/H$<2$. Using multi wavelength photometry, we have been able to obtain photometric redshifts for the hard sources and find that 10 have $0.4\le z_{\rm phot} \le 3.4$. The limiting $0.5-7$ keV flux that we reach is $\approx3\times10^{-15}$~\ergpspsqcm. Though it is impossible to constrain column-densities with the few counts in most sources, the S/H X-ray count ratios are again consistent with highly obscured AGN. 

\subsection{The Compton-thick source 15}

One source in particular, \#15 in our sample, is a red source with $B-K=6.4$. The photometric redshift implied by using detections in the \u, \b, $R$, \i, \j, \h\ and \k\ bands is \zp=0.56 with [0.4, 0.6] 90\% confidence limits. This source has been previously reported to be consistent with \zs=0.536 (\cite{lavery}), though this is a tentative estimate only.

\begin{figure}[ht]
  \begin{center}
    \epsfig{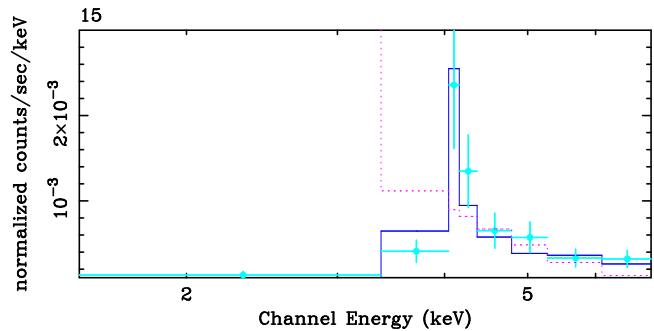}
  \end{center}
\caption{ \c\ ACIS-S3 spectrum of source \#15 in the field of A\,963. With 90 counts in all, there are 10 counts per bin. The data (blue circles with 1$\sigma$ errors) has been fit to a power-law transmission model (dark blue line) with $\Gamma=2$ and a 6.4 keV Fe K$\alpha$ line at $z=0.5$. The fitted intrinsic \nh$=1.1\times 10^{24}$ cm$^{-2}$. The pink dotted line is the power-law model affected only by Galactic line-of-sight obscuration, clearly showing the hard nature of the data.} 
\label{fig:src15xray}
\end{figure}

The X-ray source is extremely hard (Fig~\ref{fig:src15xray}), with very few counts in the soft ($0.5-2$ keV) band. With 90 counts in the whole band, we were able to extract a spectrum and fit only a crude model to the data. However, it is clear that the spectrum is consistent with a very highly obscured source. For a $\Gamma=2$ transmission model at $z=0.5$, the implied intrinsic column-density is $1.1 [0.6, 1.6] \times 10^{24}$ cm$^{-2}$ (numbers in brackets denote 90\% confidence interval), while the intrinsic L$_{2-10}=1.0\times10^{45}$~\ergps. On the other hand, an intrinsically harder spectrum with $\Gamma=1.4$ implies \nh=$8 [4, 13] \times 10^{23}$~cm$^{-2}$ and L$_{2-10}=4.6\times10^{44}$~\ergps. This source is thus a Type~II QSO discovered in X-rays. Even if $z=0$, \nh$=2\times 10^{23}$ cm$^{-2}$. Note, however, that there is a narrow peak in the observed counts at $\approx 4.1$~keV -- the energy where an Fe K$\alpha$ line redshifted from $z=0.54$ is expected to lie -- thus increasing confidence in the redshift determination.

\subsection{Clustering of serendipitous sources?}

\begin{figure}
  \begin{center}
    \epsfig{file=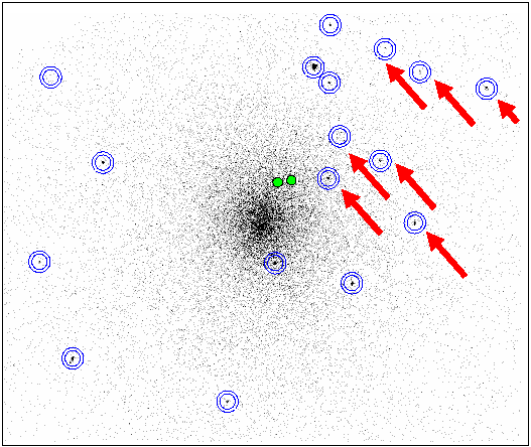, width=7cm, angle=0}
  \end{center}
\caption{ \c\ ACIS-S3 image of the field of A\,963. The cluster emission is seen close to the centre and many serendipitous sources are circled in blue. The sources with $B-K>5$ ($>7$ in one case) are marked with red arrows. North is up and east is to the left. The field of view is roughly 5.5~arcmin on a side. The green filled circles show the approximate positions of two EROs found by Smith et al. (2001) to the north-west of the cD (see text).}
\label{fig:a963clustering}
\end{figure}

To the north-west of the cluster core in A\,963, there is a group of at least 7 sources (Fig~\ref{fig:a963clustering}; these are marked with red arrows) with $B-K>5$, including an extremely red source with $B-K>7$. Three sources are detected only weakly at the limit of our \k-band UKIRT sample. Such similar colours / limits are indicative of clustering. In fact, the region which contains these 7 sources measures $\approx2$~arcmin on a side. Within such a region, only $\sim2$ sources are expected above a flux level $\approx 4\times 10^{-15}$~\ergps\ from the log$N$--log$S$ measurements (eg, \cite{mushotzky}). In addition, there are other serendipitous X-ray sources nearby (though not red in optical/NIR colour), implying an over-density. Their positions with respect to the cluster suggest that they could be strongly lensed by a factor of $\approx 20\%$.

There is yet another clue which indicates clustering of sources in this region.
\cite*{smith963eros} have recently carried out a survey of extremely-red objects (EROs) in the fields of massive clusters of galaxies, including A\,963. They found a number of EROs in the vicinity of the cD, including two to the north-west (shown as the green filled circles in Fig~\ref{fig:a963clustering}), having similar $R-K$ and $J-K$ colours which support the idea that they are both high redshift elliptical galaxies (Smith et al. note, however, that the \i-band magnitudes differ). Though these EROs are not X-ray sources, their proximity to our \lq group\rq\ of \c\ sources lends further credence to the hypothesis of clustering of \c\ sources, possibly at high redshift. With $B>25$, it will be difficult to obtain optical spectra of many of these sources.


\section{Near-IR spectra of hard serendipitous sources}

Synthesis models for the HXRB based on the absorbed AGN hypothesis
predict obscured accretion activity to peak at $z\sim 2$. At these redshifts, most
of the stronger characteristic AGN emission lines would be redshifted
into the near-IR. 
\ha\ should be detected in the \j, \h\ or \k-band at
redshifts $0.6<z<2.8$, \hb\ and the \oiii\l\l4959,5007 doublet at
$1.2<z<5$ and \pab\ at $0<z<0.9$. In addition, the 4000\AA\ continuum
break in galaxies dominated by stellar populations at least $\sim 1$ Gyr old should be observable in the near-IR
for $1.7<z<5.2$. Thus we
decided to attempt broad-band near-IR spectroscopy of the hard, optically-faint sources in our sample. Using the 4-m
United Kingdom Infrared Telescope (UKIRT), we observed five X-ray sources with the Cooled Grating Spectrograph (CGS4) during February 2000 and found that the redshifts could not be identified; all had flat and featureless spectra (\cite{c01a}).
Our equivalent-width limits of $\sim 160$\AA\ in the \k-band were improved with further UKIRT spectra obtained in May 2001 and the essential results have been confirmed, with 4 featureless sources, 1 source with single weak features in both \h\ and \k, and 3 sources with narrow multiple lines which do not fit any redshift pattern easily.

We have recently obtained deeper observations of a small sample of
optically-faint, hard X-ray sources, taken with the near-IR imaging
spectrograph ISAAC on the Very Large Telescope (VLT) in June 2001. We are able to detect clear emission lines in 3 of 8 sources, with the rest being featureless (Gandhi, Crawford \& Fabian 2002, in preparation).

\begin{figure}[ht]
  \begin{center}
    \epsfig{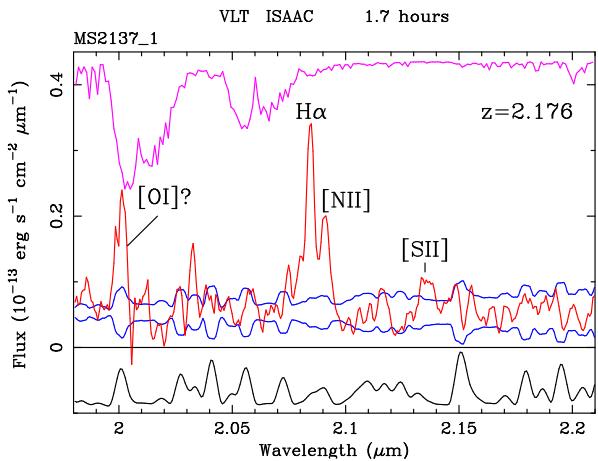}
  \end{center}
\caption{ISAAC \k-band spectrum (red) of source 1 in the field
  of the MS2137.3-2353 cluster, detailing the region of the spectrum with the
  line emission features. The 1$\sigma$ errors due to the sky are
  shown (blue lines) above and below a constant fitted continuum. The total exposure time is 102 mins for this $K=19.5$ source and the spectrum has been smoothed over 5 adjacent  pixels. The pink line above the spectrum is an arbitrarily rescaled spectrum of sky transmission, showing that any potential \oi\l6300\AA\ emission lies under strong sky absorption and overlaps with a sky OH emission feature at 2\micron\ (black line below 0).}
\label{fig:ms21371}
\end{figure}

The \k-band spectrum of our clearest redshift identification is shown in Fig~\ref{fig:ms21371}. This lies in the field of the \object{cluster MS2137.3-2353} (34.7 ks ACIS-S exposure; PI: M. Wise) and shows a clear redshifted \ha+\nii\ complex implying $z=2.176$. We detect weak \sii\l6717,6731 at the expected redshifted wavelength. There is also excess emission at the expected wavelength of redshifted \oi\l6300, but this overlaps exactly with a night sky OH emission line at $\approx 2$\micron\ and lies under a strong telluric absorption feature; since these spectra have been divided through by a spectrum of a standard star observed on the same night in order to remove telluric features, even a small difference in airmass between the target and standard star can result in incorrectly subtracted night sky lines, especially under strong absorption. Thus, it is impossible to disentangle any real \oi\l6300 emission. The X-ray S/H ratio for this source is 1.1, consistent with \nh$=2\times 10^{23}$~cm$^{-2}$ obscuring a $\Gamma=2$ power-law at $z\approx 2$. Correction for this column would imply a maximum intrinsic L$_{2-10}=2\times 10^{44}$~\ergps -- a luminous Seyfert II.

Other redshift identifications are not so clear. 4 sources are featureless within the sky poisson noise variations. 2 sources have single strong emission features, and we use the photometric redshift estimates as a first guide to their line identification: the first source in the field of \object{cluster A\,1835} has 1 strong and 1 weak emission line, which we associate with He~I~\l10830 and H~I~Pa$\delta$\l10049 at $z=1.256$. In the other source (also in the field of A\,1835), we detect a single broad line and possible \oiii\l5007, implying $z=3.831$.

The equivalent-width upper-limits that we derive for the non-detection of features range from 20--60\AA\ in the observed frame, and these would decrease further by $(1+z)$ in the rest-frame. 

Concerning the lack of significant spectral features, it is of course possible
that our bandpasses sample spectral regions which miss the emission
lines -- even with the broad \j\ and \k\ bandpasses, we are sampling only
about one-third of the optical / near-IR regime. And our equivalent-width limits are too shallow (though just marginally) to allow us to detect typical absorption lines.  

However, with our consistent campaign of searching for NIR emission features with 4 and 8 m class telescopes yielding only a single unambiguous redshift identification, it seems clear that there is a mechanism inhibiting line photon generation and/or radiation. \cite*{netzerlaor} have explored a model whereby dust in narrow line region (NLR) clouds could deplete line emission by scattering out photoionizing radiation and line photons themselves. Evidence for such dusty environments was found for at least one Type II QSO in the field of A\,2390 by \cite*{wfg} through radiative transfer modelling, as described in section 2, based on photometric detections in the {\sl ISOCAM} mid-IR bands (cf. \cite{c01b}). 

The only clear and strong forbidden line that we observe in the sample of spectra is the redshifted \nii\l6584 in one source (Fig~\ref{fig:ms21371}). Such a paucity of forbidden lines may be consistent with NLR densities greater than typical critical densities (eg, $10^6$~cm$^{-3}$ for \oiii). If this emission is spread over a parsec around the central nucleus, the implied column density is $\sim 10^{24}$ cm$^{-2}$. 

There is also the recent intriguing result that a large fraction of XRB sources found by \c\ and \xmm\ lie at $z\ltsim 1$ (eg, \cite{rosatietal}). If this is confirmed, it may may explain our lack of strong \k-band emission features, since at $z\approx 1$ the only strong emission line expected in the \k-band is \pab\ (and possibly the weaker He~I~\l10830 line).

We hope to sample other wavelength regimes for some of these sources. For example, though optical emission lines may be similarly depleted, the redshifted 4000\AA\ galaxy continuum break should be detectable, especially in the harder sources.


\section{Properties of the whole population}

\begin{figure}[ht]
  \begin{center}
    \epsfig{file=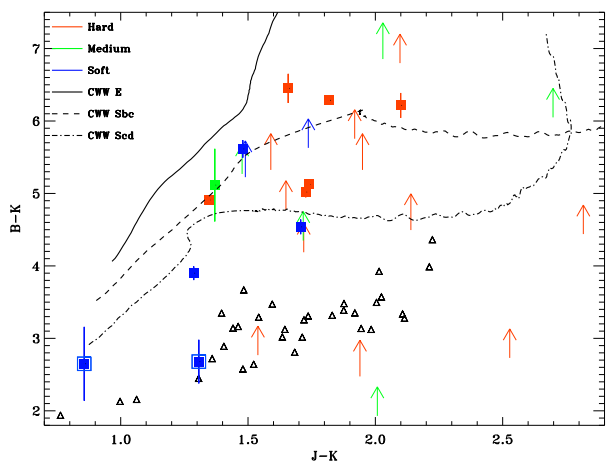, width=6.5cm, angle=90}
  \end{center}
\caption{$B$--$K$ vs. $J$--$K$ colours and limits for a subset of our sample (filled square and arrows). The small open triangle symbols mark the colours of unobscured quasars from Elvis et al. (1994) for comparison. All colours have been corrected for Galactic reddening, which is $\approx$0.1--0.5 magnitudes in $B-K$. Colour tracks are shown 
for unreddened Coleman-Wu-Weedman E (solid), Sbc (dashed) and Scd (dot-dashed) empirical templates. The redshifts of the templates increase from 0 at the left (blue end) of each line to about 0.6, 2.9 and 4 at the right for the respective templates. The red colours denote the hardest sources with S/H$<$1.5; green denotes 1.5$<$S/H$<$2.5; blue the softest with S/H$>$2.5.}
\label{fig:optirsh}
\end{figure}

Comparison with the optical--infrared
colours in Fig~\ref{fig:optirsh} shows that most of the sources have
colours or limits which are redder than unobscured AGN (\cite{elvis}; their \lq UVSX\rq\ sample). The triangles represent local ($z_{\rm median}=0.15$) bright X-ray
QSOs with little obscuration to their UV light. The relatively large $B-K$ colours of most \c\ sources  suggest
that any AGN component does not dominate the optical light, and the
colours are closer to those predicted for \cite*{cww} templates (shown without reddening) with $k$-corrections only.
The sources with the bluest $B-K$ limits have only relatively shallow 
optical lower-limits from the DSS; and the two bluest sources in the outlined squares are the broad-line quasars in the field of A\,2390 (see section 2). 

In fact, the hardest X-ray sources (red arrows and boxes) are, on average, the sources with the reddest optical-IR colour, again suggesting that in the highly obscured sources, we are viewing the host galaxy and not the AGN.

\begin{figure}[ht]
  \begin{center}
    \epsfig{file=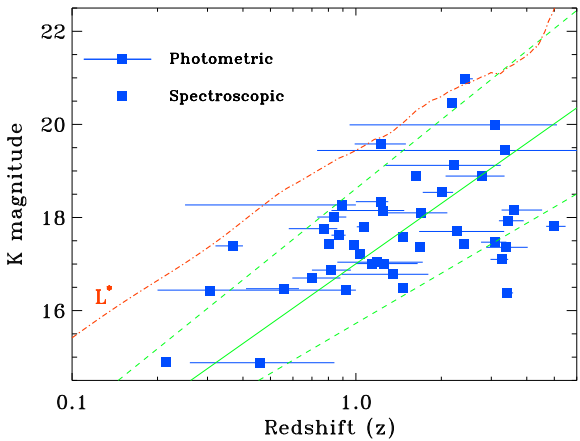, width=6.5cm, angle=90}
  \end{center}
\caption{\k-magnitude vs. Redshift for a sample of our objects from a number of cluster fields. The 90\% confidence interval is shown for objects with only a photometric estimate. The solid line shows the \k-z relation for massive radio ellipticals from Eales et al. (1993), and the dashed lines indicate the scatter of massive ellipticals about this line. The red dot-dashed line shows the predicted magnitude vs. redshift for a galaxy formed in a single stellar burst at z=10 and with a luminosity typical of an L$^*$ galaxy.}
\label{fig:kz}
\end{figure}

Fig~\ref{fig:kz} shows that we are beginning to fill up the brighter regime of the $K-z$ sample space. In fact, most of our sources are consistent with having luminosities brighter (even as much as 4 magnitudes) than L$^*$ at all redshifts. The overlap with the regime occupied by massive radio ellipticals suggests that the black holes which dominate the flux-contribution to the hard X-ray background lie in some of the most massive early-type host galaxies, in agreement with the correlation which has been found recently between host-galaxy luminosities (i.e. masses) and the masses of their central black holes (cf. \cite{magorrian}; \cite{merrittferrarese}). \cite*{f99} and other authors have attempted to model the growth of supermassive black holes through the accretion of hot gas in young, dense stellar spheroids and conclude that such black hole growth can happen in conjunction with the growth of the surrounding host and be consistent with the empirical correlations referred to above. All supermassive black holes may thus undergo an \lq XRB-emitting phase\rq\ where the intense accretion results in rapid growth as well as the radiation that we observe (a large fraction $\sim 85\%$ of which is absorbed; \cite{fi}).

\begin{figure}[ht]
  \begin{center}
    \epsfig{file=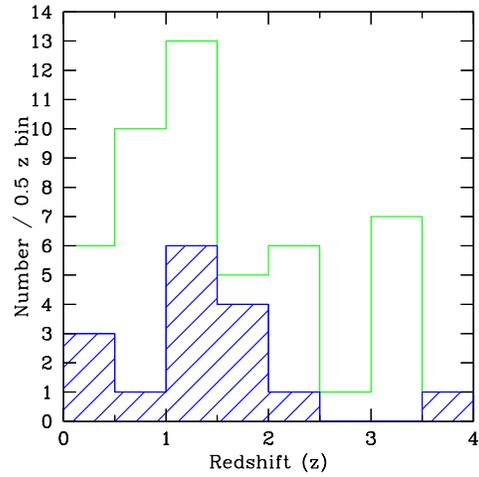, width=6.5cm, angle=0}
  \end{center}
\caption{The distribution of redshifts obtained for a sample of sources from a number of cluster fields. The shaded region represents the sources with a spectroscopic redshift measurement, the rest being photometric estimates.}
\label{fig:zhist}
\end{figure}

As discussed, recent follow-up of deep fields (eg, \cite{rosatietal}) has suggested that, contrary to expectations based on XRB synthesis models, the bulk of the X-ray source population may lie at $z\ltsim 1$. Fig~\ref{fig:zhist} shows the distribution of redshifts for our sample sources. While the distribution does peak at $z\approx 1$, we also find a tail of sources extending to higher redshifts. We hope to obtain spectra of the sources with only photometric estimates in order to place this result on a surer footing, though we note that in the field of A\,2390, the photometric redshift estimate agreed well with the spectroscopic measurement for 4 sources out of 8 using only galaxy templates. With the addition of an AGN template, the redshifts agreed in 2 more sources. Of the remaining two sources, 1 is the blue, broad-line quasar with no spectral breaks and an anomalous dust:gas ratio, while the other is a source with a close neighbour which contaminates the broad-band optical photometry obtained in non-photometric conditions. Thus, we expect the photometric redshifts to be a good first guess of the true value, especially in the hard, red sources where the optical light is dominated by the host galaxy and not the AGN.
 
\section{Conclusions}

Through a dedicated campaign of targeting the hardest and brightest sources in \c\ cluster fields, we have been able to detect 2 Type II QSOs (with L$_{2-10}>10^{45}$~\ergps\ and \nh$>10^{23}$~cm$^{-2}$. In one case, this column is probably as high as $10^{24}$~cm$^{-2}$. Both these sources show indications of strong Fe K$\alpha$ lines, though we cannot constrain the line parameters with the current data. Through radiative-transfer modelling based on {\sl ISOCAM} detections of another source, we infer the presence of warm-to-hot dust obscuring a central AGN with L$_{\rm UV}>10^{45}$~\ergps, being reprocessed into the mid-to-far infrared. We detect at least one source with an anomalous dust:gas ratio. Most sources are consistent with Compton-thin Seyfert IIs, with the redshift distribution (both spectroscopic measurements and photometric estimates) peaking at $z\approx 1$ and a significant tail to higher redshifts. These sources are optically-faint or undetected but are easily visible in the near-infrared. Detailed classification of the sources with near-infrared spectra is difficult since many sources have featureless continua, consistent with the line photons being obscured / destroyed by large column density gas and the associated dust. The \c\ sources are harboured in massive evolved host galaxies, as is expected from the bulge / black-hole inter-related growth which has been inferred in recent years. Thus, through observations of the dominant flux contributors to the X-ray background, we are beginning to find some of the truly exotic accretion sources in the distant Universe.

\begin{acknowledgements}

We thank the organizers for an enjoyable conference. Steve Allen was the PI for the A\,963 \c\ observations. PG would like to thank the Isaac Newton Trust and the Overseas Research Trust for support. ACF and CSC acknowledge financial support from the Royal Society. 

\end{acknowledgements}

\end{document}